\documentclass[twocolumn, prd, amssymb]{revtex4}
\def\CH{{\cal H}}

\def\CF{{\cal F}}

\def\IZ{\relax\ifmmode\mathchoice
{\hbox{\cmss Z\kern-.4em Z}}{\hbox{\cmss Z\kern-.4em Z}}
{\lower.9pt\hbox{\cmsss Z\kern-.4em Z}}
{\lower1.2pt\hbox{\cmsss Z\kern-.4em Z}}\else{\cmss Z\kern-.4em }\fi}
\def\IC{\relax\hbox{$\inbar\kern-.3em{\rm C}$}}
\def\IR{\relax{\rm I\kern-.18em R}}
\def\bZ{{\bf Z}}

\def\bT{{\bf T}}

\def\b{\beta}

\def\p{\pi}

\def\l{{\lambda}}
\def\o{{\omega}}
\def\O{{\Omega}}

\def\L{I}

\def\1{\relax 1 { \rm \kern-.35em I}}

\def\frac#1#2{{#1 \over #2}}

\def\p+{{\partial_+}}

\def\half{{1 \over 2}}

\def\Re{{\rm Re}}
\def\Im{{\rm Im}}

\def\[{\left [}
\def\]{\right ]}
\def\({\left (}
\def\){\right )}

\def\CH{{\cal H}}

\def\CF{{\cal F}}

\def\IZ{\relax\ifmmode\mathchoice
{\hbox{\cmss Z\kern-.4em Z}}{\hbox{\cmss Z\kern-.4em Z}}
{\lower.9pt\hbox{\cmsss Z\kern-.4em Z}} {\lower1.2pt\hbox{\cmsss
Z\kern-.4em Z}}\else{\cmss Z\kern-.4em }\fi}
\def\IC{\relax\hbox{$\inbar\kern-.3em{\rm C}$}}
\def\IR{\relax{\rm I\kern-.18em R}}
\def\bZ{{\bf Z}}

\def\bT{{\bf T}}
\def\bS{{\bf S}}
\def\bK{{\bf K}}

\def\ajou#1&#2(#3){\ \sl#1\bf#2\rm(19#3)}

\def\TrH#1{ {\raise -.5em
                      \hbox{$\buildrel {\textstyle  {\rm Tr } }\over
{\scriptscriptstyle \CH _ {#1}}$}~}}

\begin{document}

\preprint{TIFR/TH/04-23}
\preprint{SU-ITP-04-36}
%\preprint{hep-th/0409148}
\title{Exact Counting of Black Hole Microstates}
\author{Atish Dabholkar$^{a,b,c}$}
\email{atish@tifr.res.in} \affiliation{$^a$Department of
Theoretical Physics, Tata Institute of Fundamental Research,
Mumbai 400005, India\\
$^b$ Institute for Theoretical Physics,
Department of Physics, Stanford University, Stanford, CA 94305, USA\\
$^c$ Stanford Linear Accelerator Center, Stanford University,
Stanford, CA 94025, USA}

\bigskip

\date{September 2004}

\begin{abstract}
The exact entropy of two-charge supersymmetric black holes in
$N=4$ string theories is computed to all orders using Wald's
formula and the supersymmetric attractor equations with an
effective action that includes the relevant higher curvature
terms. Classically, these black holes have zero area but the
attractor equations are still applicable at the quantum level. The
quantum corrected macroscopic entropy agrees precisely with the
microscopic counting for an infinite tower of fundamental string
states to all orders in an asymptotic expansion.
\end{abstract}

\maketitle

%\centerline{\it Introduction}
%\medskip

A distinctive feature of superstring theory is that its spectrum
often contains an infinite tower of BPS states in a given
topological sector. The first example of such a tower of BPS
states was noticed in the perturbative spectrum of toroidally
compactified superstring theories
\cite{Dabholkar:1989jt,Dabholkar:1990yf}. We will be interested
here in the heterotic string  compactified on $\bT^4 \times \bT^2$
where $\bT^4$ is a 4-torus in $\{6789\}$ directions and $\bT^2$ is
a 2-torus which we take to be a product of two circles in the
$\{45\}$ directions. Consider now a string state with winding
number $w$ along the $x_5$ direction. In a given winding sector,
there is a tower of BPS states each in the right-moving ground
state but carrying arbitrary left-moving oscillations subject to
the Virasoro constraint $ N_L = 1 -nw$,
%\begin{equation}\label{Virasoro}
%    N_L = 1 -nw,
%\end{equation}
where $N_L$ is the left-moving oscillation number and  $n$ is the
quantized momentum along $x_5$
\cite{Dabholkar:1989jt,Dabholkar:1990yf}. Note that $N_L$ is
positive and hence a BPS state that satisfies this constraint has
negative $n$ for positive $w$ for large $N_L$. We will henceforth
denote these states by $(n, w)$.

The number of such states is summarized conveniently by a
partition function
\begin{equation}\label{partition}
    Z(\beta) = 16  \sum d_N e^{-\b N},
\end{equation}
where $N \equiv w|n| = N_L -1$. The factor of $16$ comes from the
degeneracy of the right-moving supersymmetric ground state. Since
$N_L$ is the number operator for the $24$ left-moving transverse
bosons, the partition function can be readily evaluated
\begin{equation}\label{partition2}
    Z(\b) = {16\over \Delta(q)},
\end{equation}
where $\Delta(q)$ is the Jacobi discriminant function with
argument $q =\exp{(-\b)}$. In terms of the Dedekind eta function
$\eta(q)$, the discriminant is given by $\Delta(q) =
\eta(q)^{24}$.

The number of states at level $N$  is then given by the inverse
Laplace transform
\begin{equation}\label{density}
    d_N = {1\over 2\pi i} \int_C d\b e^{\b N}  {1\over
    \Delta(e^{-\b})}.
%    = {1\over 2\pi i} \oint {dq \over q^{N+1}} {1\over \Delta(q)}.
\end{equation}
To find the asymptotic density at large $N$, we want to take the
high temperature limit or $\beta \rightarrow 0$. It is convenient
to use the modular property of the discriminant
\begin{equation}\label{modular}
    \Delta(e^{-\b}) = ({\b\over 2\pi})^{-12}
    \Delta(e^{-4 \pi^2 /\b}).
\end{equation}
As $e^{-4 \pi^2 /\b} \rightarrow 0$, we can  then use the
asymptotics $ \Delta(q) \sim q$ and evaluate the integral
\begin{equation}\label{asymptotic}
   d_N = {1\over 2\pi i} \int_C d\b {e^{\b N}}({\b\over 2\pi})^{12}
    {1\over \Delta(e^{-4 \pi^2 /\b})}
\end{equation}
in saddle-point approximation. The saddle point occurs at $\beta =
2\pi / \sqrt{N}$ and the degeneracy has the characteristic
exponential growth $ d_N  \sim \exp{( 4 \pi \sqrt{w|n|})}$. The
sub-leading terms can be computed in an asymptotic expansion.

This tower of states has played a crucial role in furthering our
understanding of dualities and  black hole physics. Heterotic
string on $\bT^4 \times \bT^2$ is dual to Type-IIA on $\bK_3
\times \bT^2$ \cite{Hull:1994ys,Witten:1995ex}. Initial evidence
for this duality came from matching the low-lying BPS states and
the supergravity action but a far more stringent test is obtained
by matching the entire infinite tower of BPS states. The state
$(n, w)$ is dual to $w$ NS5-branes wrapping $\bK_3 \times \bS^1$
carrying $n$ units of momentum which in turn is dual to $w$
D4-branes wrapping the $\bK_3$ with a gas $n$ D0-branes on its
worldvolume. The characteristic exponential growth of such brane
states was computed in \cite{Vafa:1995zh}. In fact, this partition
function makes its appearance also in topologically twisted Yang
Mills theories on $\bK_3$ which provided one of the early hints of
the stringy duality \cite{Vafa:1994tf}. We will return to this
dual description subsequently.

Another important application of this tower of states comes from
its relation to black hole entropy. The state $(n, w)$ corresponds
to a charged extremal black hole in four dimensions and one would
expect that the logarithm of $d_N$ for large $N$  should match the
Bekenstein-Hawking entropy of this black hole. The corresponding
black hole solutions were obtained in \cite{Sen:1994eb} which can
also be seen to arise directly from the dimensional reduction of
the underlying winding string solution in five dimensions
\cite{Dabholkar:1995nc}. Classically, these black holes have zero
area and would appear to have zero entropy but the higher
curvature corrections to the supergravity action can correct the
solution. Assuming that the string corrections result in a finite
area horizon it was shown in \cite{Sen:1995in} that the black hole
then has nonzero entropy in agreement with the logarithm of the
degeneracy. In particular, the nontrivial functional dependence
$\sqrt{n|w|}$ on the charges is correctly reproduced. However, the
precise numerical factor of $4\pi$ cannot be computed and the
assumption of finite area is in need of further evidence.
Subsequent developments have focused mostly on black holes
carrying three or more charges that have finite horizon area
already classically \cite{Strominger:1996sh} so that the precise
numerical factor can be computed reliably within the supergravity
approximation.

In this note we return to the  two charge black holes
corresponding to the states $(n, w)$ and show that it is possible
to take into account exactly all  higher curvature corrections to
the entropy. After incorporating these corrections and using the
exact entropy formula due to Wald
\cite{Wald:1993nt,Iyer:1994ys,Jacobson:1994qe} for the fully
corrected action, the precise numerical factor $4\pi$ in
(\ref{asymptotic}) can be computed. One can even go further and in
fact reproduce not only the leading exponential but the entire
asymptotic expansion of the partition function (\ref{partition2})
exactly to all orders for large $N$.

Before proceeding further let us note the following curious fact.
The curvature squared coupling in the four-dimensional effective
action for the heterotic string on $\bT^6$ is of the form
\begin{equation}\label{hetcoupling}
      {1\over 16 \pi} {\rm Re} [\int {\log{\Delta(q)} \over 2\pi i}\,
       {\rm tr}(R-i R^*)^2],
\end{equation}
with $q= e^{2\pi i \l}$ where $\l$ is the dilaton-axion field, $R$
is the curvature 2-form and the trace is in the tangent space
$SO(1,3)$ representation \cite{Harvey:1996ir}. The action with its
coefficient can be deduced easily from S-duality and Green-Schwarz
mechanism. What is striking is that the Jacobi discriminant
function that we encountered in (\ref{partition2}) makes its
appearance here in a completely different context. At first sight
this would appear to be little more than a coincidence. After all,
the argument $q$ in (\ref{hetcoupling}) depends on a spacetime
field whereas in (\ref{partition2}) it depends on the worldsheet
temperature $\beta$. The q-expansion in (\ref{hetcoupling}) gives
the nonperturbative corrections to the effective coupling from
5-brane instantons \cite{Harvey:1996ir} whereas the q-expansion in
(\ref{partition2}) counts the spectrum of perturbative BPS winding
states.  It turns out, however, that there is indeed a deep and
precise connection between the two that is provided by the
supersymmetric attractor mechanism
\cite{Ferrara:1995ih,Strominger:1996kf,Ferrara:1996dd}, its
elegant implementation in supergravity using Wald's formula for
higher derivative F-term corrections pioneered in
\cite{LopesCardoso:1998wt,LopesCardoso:1999cv,
LopesCardoso:1999xn,LopesCardoso:2000qm,Maldacena:1997de}, and the
recent proposal for the black hole partition function in
\cite{Ooguri:2004zv}.

Let us summarize the relevant formalism
\cite{LopesCardoso:1998wt,LopesCardoso:1999cv,
LopesCardoso:1999xn,LopesCardoso:2000qm,Ooguri:2004zv}. To be
closer to the discussion in the literature, we work in the dual
description of Type-IIA on $\bK_3 \times \bT^2$ which can be
viewed as a special case of a Calabi-Yau 3-fold. The resulting
supergravity in four dimensions has $N=4$ supersymmetry but it
will be convenient for our purposes to use the $N=2$ notation of
special geometry.

The vector multiplet moduli space of $N=2$ supergravity with $n_v$
vector multiplets is parameterized by $n_v +1$ complex projective
coordinates $X^\L, \L = 0, 1, \ldots, n_v$. There are an infinite
number of higher derivative corrections to the Einstein-Hilbert
action that are expected to be relevant for the computation of the
entropy. These F-type corrections to the effective action are
summarized by the string-loop corrected holomorphic prepotential
\begin{equation}\label{prepotential}
F(X^\L,W^2)= \sum_{h=0}^\infty F_h(X^\L)W^{2h},
\end{equation}
where $F_h$ are computed by the topological string amplitudes
\cite{Bershadsky:1993cx,Bershadsky:1993ta,Antoniadis:1993ze,
Gopakumar:1998ii,Gopakumar:1998jq} and $W^2$ is the reduced chiral
multiplet \cite{LopesCardoso:1999cv,LopesCardoso:1999ur} that
involves the graviphoton field strength. The prepotential obeys
the homogeneity relation
\begin{equation}\label{homogeneous}
    X^\L\partial_\L F(X^\L,W^2)+W\partial_{W} F(X^\L,W^2)=2 F(X^\L,W^2).
\end{equation}
The moduli couple to the electromagnetic fields and as a result
vary with the radius in the back hole background. Starting with
arbitrary values at infinity, at the horizon they approach an
attractor point in the moduli space. The values at the attractor
point are determined by the black hole attractor equations
\begin{equation}\label{pattractor1}
   p^\L=\Re[CX^\L],
\end{equation}
\begin{equation}\label{qattractor1}
    q_\L=\Re\left[CF_\L\left(X^\L, {256 \over C^{2}}\right),
\right]
\end{equation}
and $C^2W^2=256$, where $F_\L \equiv {\partial{F}
/\partial{X^\L}}$ are the holomorphic periods. The scaling field
$C$ is introduced so that $(CX^\L, CF_\L)$ is non-projective and
transforms like $(p^\L, q_\L)$ as a vector under the $Sp(2n_v +2;
\bZ)$ symplectic duality group. The attractor equations are then
determined essentially by symplectic invariance. For a recent
review of the leading order attractor equations and their
applications see \cite{Moore:2004fg}.

The quantum corrected black hole entropy is given by
\cite{LopesCardoso:1998wt,LopesCardoso:1999cv,
LopesCardoso:1999xn,LopesCardoso:2000qm}
\begin{equation}\label{entropy1}
S_{\rm BH}={\pi i \over 2}(q_{\L}\bar C\bar X^\L-p^\L \bar C\bar
F_{\L})+{\pi \over 2}{\Im}[C^3\partial_C F].
\end{equation}
The first set of attractor equations (\ref{pattractor1}) can be
solved by
\begin{equation}\label{pattractor2}
C X^I = p^I + {i \over \pi} \phi^I
\end{equation}
in terms of the `potentials' $\phi^\L$. Then the entropy
(\ref{entropy1}) can be written in a suggestive form
\cite{Ooguri:2004zv} as
\begin{equation}\label{entropy2}
S_{\rm BH}(q , p) = {\cal F}(\phi , p ) - \phi^\L {\partial \over
\partial \phi^\L} {\cal F}(\phi, p).
\end{equation}
in terms of a `free energy' function
\begin{equation}\label{free}
{\cal F}(\phi , p ) = - \pi {\rm Im} \left[ F\left( p^I + {i \over
\pi} \phi^I, 256\right) \right].
\end{equation}
The potentials $\phi^\L$ in this equation are determined in terms
of the charges by the second set of attractor equations
(\ref{qattractor1})
\begin{equation}\label{qattractor2}
q_\L=\half(CF_\L+ \bar C\bar F_\L) = -{\partial \over \partial
\phi^\L} {\cal F}(\phi, p).
\end{equation}
Given the form of the entropy (\ref{free}), it is natural to
define a `partition function' as suggested in \cite{Ooguri:2004zv}
\begin{equation}\label{partition3}
Z_{\rm BH}(\phi^\L,p^\L)= e^{{\cal F}(\phi^\L, p^\L)} \equiv
\sum_{q_\L}\Omega(q_\L, p^\L)e^{- \phi^\L q_\L},
\end{equation}
where $\Omega(q_\L, p^\L)$ are the black hole degeneracies. The
Boltzmann entropy $\ln{\Omega(q, p)}$ is then expected to agree
with the thermodynamic entropy $S_{\rm BH}(q , p)$ in
(\ref{entropy2}) for large charges.

It would be very interesting to test the proposal
(\ref{partition3}) for the black hole partition function by
comparing it with the microscopic partition function. For a
general Calabi-Yau compactification, such an explicit comparison
is difficult for a number of reasons. On the supergravity side, to
make this comparison it is necessary to compute all infinite terms
$F_h$ in the prepotential (\ref{prepotential}). Even though these
are given in principle by the topological string, they are not
always explicitly computable. On the microscopic side, the
counting of states is complicated by the fact that the number of
BPS states can jump in $N=2$ supersymmetric theories
\cite{Seiberg:1994rs}. This phenomenon is possibly related to
black hole fragmentation \cite{Maldacena:1998uz} and multiple
basins of attraction \cite{Moore:1998pn, Denef:2000nb} as
suggested in \cite{Ooguri:2004zv}. In addition, there are
subtleties having to do with the holomorphic anomalies and the
background dependence on hypermultiplet moduli which complicate
the picture further\cite{Ooguri:2004zv}.

One virtue of the tower of states in  $N=4$ compactification that
we have considered is that it provides a particularly simple but
nontrivial example for making a clean comparison between black
hole microstates and the exact entropy formula. For this system,
on the macroscopic side the prepotential is explicitly computable.
Moreover on the microscopic side the exact partition function of
the microstates is also known and is given by (\ref{partition}).
With $N=4$ supersymmetry we do not expect that the number of BPS
states would jump.

For Type-IIA on a Calabi-Yau manifold, in the large volume limit,
$F_0$ and $F_1$ are given by
\begin{equation}\label{classicalF}
    F_0 =  -{1\over 6} C_{IJK} {X^I X^J X^K
\over X^0}, \qquad F_1 = -{1\over 64} {c_2 \over 24} {X^1\over
X^0},
\end{equation}
where $A = 1, \ldots, n_v$ and $c_2$ is the second Chern class and
$C_{IJK}$ are the intersection numbers of a basis $\{\Sigma^I\}$
of 4-cycles \cite{LopesCardoso:1999ur}. For a properly normalized
basis of 2-forms $\{\o_I\}$ that are Poincar{\'e} dual to
$\{\Sigma^I\}$, the intersection numbers are given by
\begin{equation}\label{interesection}
C_{IJK} = \int_{CY3} \o_I \wedge \o_J \wedge \o_K.
\end{equation}
In the special case of $\bK_3 \times \bT^2$, there are $23$
2-cycles of which we take $w_1$ to be  the 2-torus itself and
$w_a, a =2, \ldots, 23$ to be the $22$ 2-cycles of $\bK_3$. The
$N=2$ reduction of $N=4$ is a bit subtle in supergravity because
of the extra gravitini multiplets. For the particular charge
configuration that we have chosen, however, the fields in the
gravitini multiplets are not excited. Hence we can safely ignore
them.

A major simplification for $\bK_3 \times \bT^2$ is that in
(\ref{prepotential}), all $F_h$ for $h >1$ vanish. This can be
seen most easily in the corresponding topological string from the
counting of fermion zero modes. Moreover, $F_0$ is given by its
classical value and receives no quantum corrections because $F_0$
determines the metric on the moduli space which is known to
receive no corrections in $N=4$ supergravity. Thus, the only
nontrivial term in the prepotential comes from $F_1$ which has
already been computed in the literature in  a number of different
ways -- either directly from its definition
\cite{Bershadsky:1993ta, Dixon:1990pc}, or by using the
holomorphic anomaly \cite{Bershadsky:1993cx}, or from
string-string duality \cite{Harvey:1996ir} by requiring agreement
with (\ref{hetcoupling}). The fully quantum corrected prepotential
then takes a particularly simple form
\begin{equation}\label{prepotential2}
F(X, W^2)= -\half C_{ab} X^a X^b ({X^1 \over X^0}) - {W^2 \over
128 \pi i}\log{\Delta{(q) }}
\end{equation}
where $q = \exp{(2\pi i X^1 /X^0)}$ and $C_{ab}$ is the
intersection matrix of $K3$ and we have used the fact that $c_2
=24$ for $\bK_3$. It can be seen using the action in
\cite{LopesCardoso:1998wt, deWit:1996ix} that $X^1/ X^0$ is the
correctly normalized dilaton-axion field $\l$ in the heterotic
description that couples to integral Pontryagin class so that the
action is then invariant under $\l \rightarrow \l +1$.

Now that we have the exact prepotential, let us see  what our
charge configuration looks like in this basis. The perturbative
state $(n, w)$ on the heterotic side is dual to $w$ 4-branes
wrapping the $\bK_3$ with $n$ 0-branes sprinkled on it. The
4-cycle is dual to the 2-form $w_1$ and hence we have a nonzero
magnetic charge $p_1 =w$ and all other magnetic charges are zero.
The 0-brane couples electrically to the graviphoton field and
hence $q_0 = n$ and all other electric charges are zero. We can
readily evaluate the free energy defined in (\ref{free})
\begin{equation}\label{free2}
    \CF(\phi, p) = -{1\over 2} C_{ab} \phi^a\phi^b {p^1 \over \phi^0}-
    \log {(|\Delta(q)|^2)}
\end{equation}
with
\begin{equation}\label{argument}
    q = \exp{({2\pi^2 p^1 \over \phi^0} + {2\pi i \phi^1 \over \phi^0})}.
\end{equation}

In the large volume limit  we can approximate the second term in
(\ref{free2}) by $-{ 4\pi}^2 p^1 /\phi^0$. It is then easy to
solve the attractor equations for the set of charges $q^A =(q^0,
0, 0, \ldots, 0)$ and with $p_A=(0, p_1, 0, \ldots, 0)$. The
solution is
\begin{equation}\label{solution}
    \phi^0 = -2\pi \sqrt{p^1\over|q_0|}, \qquad \phi^a = 0,
\end{equation}
and $\phi^1$ is undetermined. The entropy is given by
\begin{equation}\label{entropy3}
    S = 4\pi \sqrt{p^1 |q_o|} = 4\pi \sqrt{w |n|},
\end{equation}
which matches exactly with the logarithm of the degeneracy
(\ref{asymptotic}) of the tower of states at large $N$. It is
remarkable that once the higher derivative corrections are
included, the attractor formalism is powerful enough to correctly
reproduce the entropy even for black holes that have zero area
classically. This result is implicit in some of the early work
\cite{LopesCardoso:1999ur} however there the focus is on computing
the corrections to the entropy of black holes that have finite
area already classically.

Encouraged by this, we would now like to reproduce the exact
degeneracy $\O(q^A, p_A)$ in (\ref{partition3}) which is given by
the inverse Laplace transform,
\begin{eqnarray}\nonumber\label{degeneracy1}
({1\over 2\pi} )^{24} \int \prod_{a=1}^{22} d\phi^a d\phi^1
  d\phi^0 \exp{(\CF(\phi^A, p^A) + \phi^0 q_0)}.
\end{eqnarray}
Using the free energy (\ref{free2}), the Gaussian integrals over
$\{\phi^a\}$ can be performed immediately to get,
\begin{equation}\label{degeneracy2}
    \Omega(q_0, p^1) \sim {1\over 2\pi} \int dx ({x\over 2\pi})^{12}
    e^{Nx}\int {d\theta \over  2\pi} {1\over |\Delta(q)|^2},
\end{equation}
where we have defined $x = -\phi^0/p^1$ and $\theta =
\phi^1/\phi^0$ so that $q =\exp{(2\pi i \tau)}$ with $\tau =
\theta + \pi i/x $ and $N = -p^1 q_0 \equiv w|n|$. For large $N$,
since the saddle point is localized in the small $x$ region, we
can use the approximation
\begin{equation}\label{approximation}
|\Delta(q)|^2 \sim  |q|^2 \sim e^{-4\pi^2/ x}
\end{equation}
in (\ref{degeneracy2}). The $\theta$ integral can be done
trivially and the $x$ integral is then identical to the asymptotic
form of the $\beta$ integral (\ref{asymptotic}) in the region of
small $\beta$ where the saddle point is localized.
%We can use the modular property (\ref{modular}) to write the
%integral as
%\begin{equation}\label{final}
%\Omega(q_0, p^1) \sim {1\over 2\pi i} \int dx
%    e^{Nx} {1\over \Delta(e^{-x})}.
%\end{equation}
%Here we have used the approximation $ -{1\over \tau} \sim {i\over
%2 \pi x}$ for small $x$ to perform the $\theta$ integral.
Hence, up to a multiplicative normalization, the black hole
degeneracy of states (\ref{degeneracy2}) matches with the
microscopic degeneracy of states (\ref{asymptotic}) exactly to all
orders in an asymptotic expansion at large $N$. A more complete
analysis will be presented elsewhere \cite{Progress}.

%\centerline{\it Acknowledgments}

I am grateful to Bernard de Wit  for valuable discussions and
correspondence. I would also like to thank Rajesh Gopakumar for
discussions and Andy Strominger for comments on the draft. This
work was supported in part by Stanford Institute of Theoretical
Physics, the David and Lucile Packard Foundation Fellowship for
Science and Engineering, and by the Department of energy under
contract number DE-AC02-76SF00515.

\medskip
\bibliography{ref}
\bibliographystyle{apsrev}
\end{document}